\documentclass[runningheads]{llncs}

\usepackage{graphicx}
\usepackage{amsmath,amssymb,amsfonts}
\usepackage{algorithm}
\usepackage{textcomp}
\usepackage[table,xcdraw]{xcolor}
\usepackage{listings}
\definecolor{strings_color}{rgb}{1.0,0.0,0}
\definecolor{comments_color}{rgb}{0.2,0.2,0.2}
\usepackage{hyperref}
\def\code#1{\texttt{#1}}

\begin{document}
	\sloppy
\title{Posit NPB: Assessing the Precision Improvement in HPC Scientific Applications}
\titlerunning{Assessing the Precision Improvement of Posit in the NPB}
%
\author{Steven W. D. Chien\inst{1} \and
Ivy B. Peng\inst{2} \and
Stefano Markidis\inst{1}}
\authorrunning{S. W. D. Chien et al.}
%
\institute{$^1$KTH Royal Institute of Technology, Stockholm, Sweden \\ $^2$Lawrence Livermore National Laboratory, Livermore, CA, USA}

\maketitle              
\begin{abstract}
Floating-point operations can significantly impact the accuracy and performance of scientific applications on large-scale parallel systems. Recently, an emerging floating-point format called Posit has attracted attention as an alternative to the standard IEEE floating-point formats because it could enable higher precision than IEEE formats using the same number of bits. In this work, we first explored the feasibility of Posit encoding in representative HPC applications by providing a 32-bit Posit NAS Parallel Benchmark (NPB) suite. Then, we evaluate the accuracy improvement in different HPC kernels compared to the IEEE 754 format. Our results indicate that using Posit encoding achieves optimized precision, ranging from 0.6 to 1.4 decimal digit, for all tested kernels and proxy-applications. Also, we quantified the overhead of the current software implementation of Posit encoding as 4$\times$-19$\times$ that of IEEE 754 hardware implementation. Our study highlights the potential of hardware implementations of Posit to benefit a broad range of HPC applications.

\end{abstract}

\section{Introduction}\label{sec:introduction}
Floating-point operations are indispensable for many scientific applications. Their precision formats can significantly impact the power, energy consumption, memory footprint, performance, and accuracy of applications. Moving towards exascale, optimizing precision formats in HPC scientific applications could address some key challenges identified on exascale systems~\cite{dongarra2011international}. Recent works in hardware-supported half-precision, software-guided mixed-precision, and adaptive precision have highlighted the importance of reconsidering precision formats~\cite{anzt2019adaptive,markidis2018nvidia,Menon:2018:AAD:3291656.3291720}. \textit{Posit}~\cite{gustafson2017beating}, an alternative to IEEE 754 floating-point format, has gained increasing attention in the HPC community because its \textit{tapered precision} can achieve higher precision than IEEE 754 format using the same number of bits. Posit has been explored in Deep Learning applications~\cite{johnson2018rethinking}, Euler and eigenvalue solvers~\cite{Lindstrom:2018:UCR:3190339.3190344}. Still, its precision improvements in general HPC scientific applications require systematic efforts to understand and quantify, which motivates our study in this paper. Our work provides a 32-bit Posit implementation of the popular NAS Parallel Benchmark (NPB) suite~\cite{nas,bailey1991parallel}, called \textit{Posit NPB} to quantify the improved precision using Posit formats compared to 32-bit IEEE 754 format in representative HPC kernels. Our main contributions are as follows:
\begin{itemize}
\item We provide a publicly available 32-bit Posit implementation of the NPB benchmark suite
\item We define the metric for accuracy and use it to quantify the precision improvements using Posit formats in five kernels and proxy-applications compared to 32-bit IEEE 754 format
\item We also provide a 128-bit IEEE 754 floating-point (Quad) implementation of the NPB benchmark suite as a high-precision solution reference
\item Our Posit implementation exhibit 0.4 to 1.6 decimal digit precision improvement in all tested kernels and proxy-applications compared to the baseline
\item We quantified the overhead of software-based Posit implementation as 4$\times$-19$\times$ that of IEEE 754 hardware implementation
\item We show that Posit could benefit a broad range of HPC applications but requires low-overhead hardware implementation.
\end{itemize}
\section{Floating-point Formats}\label{sec:background}
Fractional real numbers are represented as floating-point numbers, and their operations are defined by floating-point operations in computer arithmetics. Instead of representing the number in its original form, a number is represented as an approximation where the trade-off between precision and range is defined. Given the same amount of memory space, a larger range of numbers can be represented if numbers in that range use a less accurate approximation.

IEEE floating-point numbers are often represented by three components, i.e., a sign bit, an exponent, and a significand. A sign bit represents whether the number is positive or negative. An exponent represents the shifting that is required to acquire the non-fraction part of a number. Finally, a significand represents the actual number after shifting. Currently, IEEE 754 format is the most broadly adopted standard.

Posit format uses four components, i.e., a sign bit, a regime, an exponent and a fraction. Different from IEEE Float, these components could have variable sizes. The first component after the sign bit is \emph{regime}, which is used to compute a scaling factor $useed^k$ where $useed=2^{2^{es}}$. The regime component encodes a number $k$ through a prefix code scheme. The regime contains several consecutive $1$s or $0$s, which is terminated if the next bit is the opposite. $k$ is defined by $k=-m$, where $m$ is the length of the bit string before the opposite bit when the bits are all zero. For instance, if the bit string is all $1$s and terminated by a $0$, $k$ is defined by $k=m-1$. After the regime, depending on the number of bits left, the \emph{exponent} begins and runs for a maximum length of $es$. The exponent encodes an unsigned integer, which represents another scaling factor $2^{exponent}$. Finally, the \emph{fraction} has the same functionality as the significand in IEEE Float.

\begin{figure}[h]
	\begin{center}
		\includegraphics[width=0.95\linewidth]{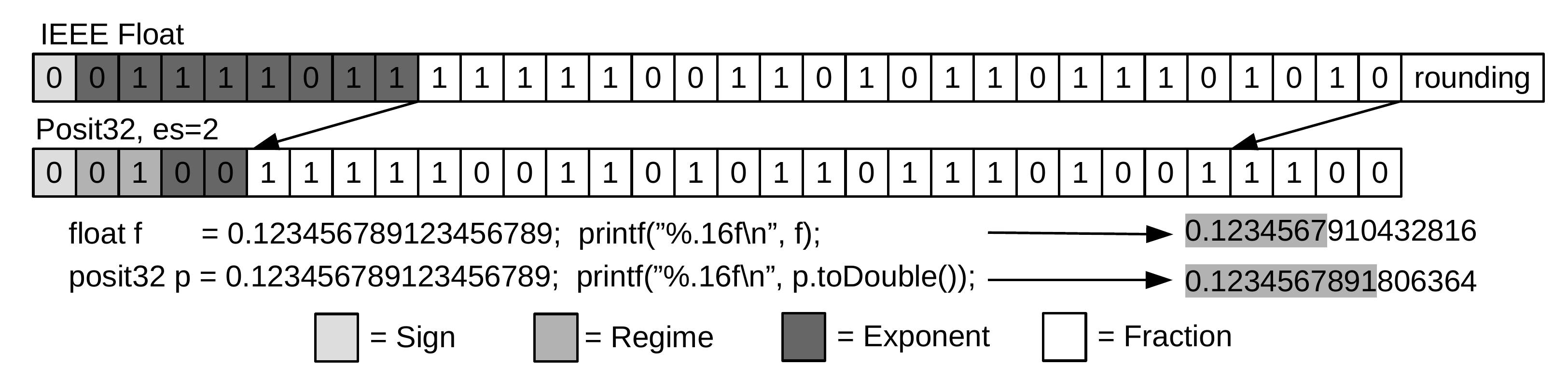}
		\caption{Binary formatting of IEEE Float and 32 bit Posit with $es=2$ when representing an arbitrary fractional number.}
		\label{fig:posit-format}
	\end{center}
\end{figure}

Fig.~\ref{fig:posit-format} illustrates a number encoded in IEEE Float and  32-bit Posit with $es=2$. The string begins with a zero that indicates the number is positive. The \emph{regime} bit runs for $01$, which means that $m=1$ and thus $k=-1$. Since $es=2$, $useed=2^{2^{es}}=16$, the scaling factor $useed^k=1/16$. After the termination of \emph{regime}, the exponent begins and has a length of $es=2$. The exponent $00$ is represented by $2^{exponent}=2^0=1$. The remaining bits are used for the fraction, which encodes $1 + 130903708/2^{27}$, where the one is implicit, and the size of the fraction is 27. Since the scheme has a smaller exponent than IEEE Float, more bits can be used in the fraction, which attributes to higher accuracy.

In this work, we also use a special type in Posit, called \emph{Quire}, to facilitate high accuracy Fused Multiply-Add (FMA). Quire can be considered as a large scratch area for performing high precision accumulation and deferred rounding during FMA. Quire requires a large number of bits. For instance, it requires 32 bits for 8-bit Posit, 128 bits for 16-bit Posit, and 512 bits for 32-bit Posit.
\section{Methodology}\label{sec:method}
Our work aims to assess precision optimization by Posit in HPC scientific applications. To achieve this, we choose a widely-adopted parallel benchmark suite on HPC systems, the NAS Parallel Benchmark (NPB) suite~\cite{bailey1991parallel}. The NPB suite was originally derived from Computational Fluid Dynamics (CFD) applications, and closely mimic real-world HPC applications. The suite includes five kernels (IS, EP, CG,  MG, and FT) and three proxy-applications (BT, LU, and SP). In this work, we extend a subset of the suite that uses floating-point operations to evaluate the impact of Posit arithmetic in typical HPC applications. We based our implementation on the C version of the suite~\cite{nas}. Our Posit NPB suite includes CG (Conjugate Gradient), MG (Multigrid), LU (Lower-Upper decomposition solver), BT (Block Tridiagonal solver) and FT (FFT solver). The benchmark suite is publicly available in a code repository\footnote{\url{https://github.com/steven-chien/NAS-Posit-benchmark}}. 

The original NPB implementation uses only the 64-bit IEEE 754 floating-point format. We provide a 32-bit Posit implementation and IEEE 754 floating-point implementation of the suite. To compare with a high accuracy IEEE format, we additionally provide a 128-bit IEEE floating point (quad) implementation. The NPB suite predefines problem sizes into different classes: S class for small tests, W class for workstation-size tests; A, B, C classes for standard test on supercomputer and E, D and F for large test problems on parallel machines. Our evaluation includes experiments using various problem classes for understanding the impact of Posit arithmetics. 

We define the metric for accuracy as the difference between the approximated value in various precision formats and the exact value. We then evaluate the accuracy using five precision formats, i.e., Quad~(\code{quad}), Double~(\code{double}), IEEE Float~(\code{float}), 32 bit Posit~(\code{posit32}) and Quire for 32~bit Posit~(\code{quire32}). For Quad precision, as it is not natively supported by C and C++, we adopted the \code{libquadmath} library by GCC. For Posit and Quire types, we used the C++ interface provided by the \emph{SoftPosit} library where operator overloading is supported. For all evaluation, we use single-thread executions of kernels to avoid interference from multiple threads. We validate the solution from each kernel in the highest accuracy, i.e., Quad precision. 
We cast the generated results to the selected precision to control error propagation resulted only from the computation other than problem generation.

\section{Results}\label{sec:results}

We evaluate our NPB extension on a workstation with an Intel Core i7-7820X CPU with eight cores and 16 threads. The system has 32GB of RAM with a 480GB SSD. The operating system is Fedora 29 running on Kernel version 4.19.10-300. The compiler used is GCC 8.2.1 and the latest SoftPosit library from main development branch is used\footnote{SoftPosit commit 688cfe5d}. We compute the machine epsilon ($\epsilon$) of different formats using linear search method for reference. On this workstation, the measured $\epsilon$ values are 1.92E-34, 2.22E-16, 1.19E-7, 7.45E-9 and for \code{quad}, \code{double}, \code{float} and \code{posit32} respectively.

Reproducing floating-point computation results across platforms and optimization is a difficult task. When compiling the benchmarks, we have used the flags \code{-ffp-contract=off} for turning off possible Fused Multiply Add (FMA) and \code{-ffloat-store} to avoid storage of floating-point variables in registers thus avoiding the effect of storing intermediate value in extended precision by the processor. We select three problem sizes for the precision evaluation: class S, W and A. 

{\bf CG: Conjugate Gradient.} The CG benchmark computes the smallest eigenvalue of a sparse matrix with the Inverse Power method. At each Inverse Power iteration, a linear system is solved with the Conjugate Gradient (CG) method. The CG kernel consists of three main steps: first, the vector $x$ is initialized randomly; second, the CG method solves the linear system $Az = x$ where $A$ is symmetric, definite-positive matrix, and $x$ in the known term of the linear system; third, the inverse power method uses the $||r||$ that is the norm of the residual vector $r$ to calculate $\zeta = \lambda + 1/x^Tz$ where $\lambda$ is the shift factor for different problem size. In our set-up, each Inverse Power iterations includes 25 CG iterations.

\begin{figure}[h]
	\begin{center}
		\includegraphics[width=\linewidth]{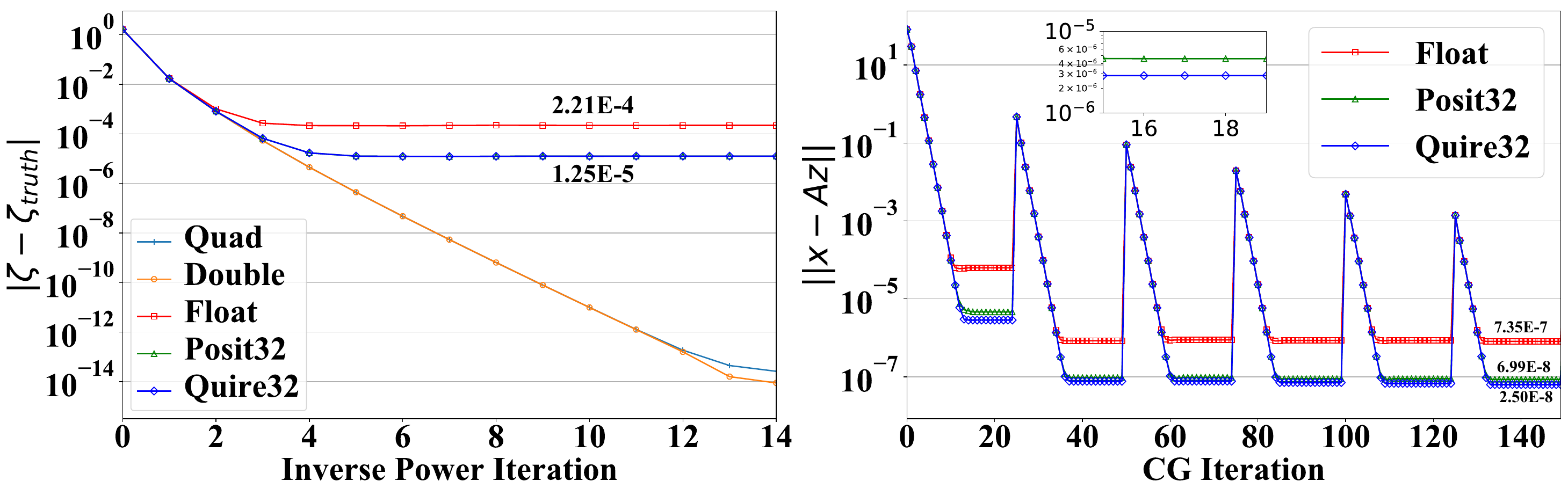}
		\caption{CG: The left panel displays the error in evaluating $\zeta$ for the first 15 Inverse Power iterations and problem class W using different floating-point formats. The error decreases until reaching a minimum value approximately after five iterations for \code{posit32} and \code{float}. The right panel displays residual error of CG iterations. At the end of each Inverse Power iteration (25 CG iterations), the norm of the residual increases by more of two orders. The zoom-in view of on CG residual shows the difference in error between \code{posit32} and \code{quire32}.}
		\label{fig:cg-W}
	\end{center}
\end{figure}

The left panel of Fig.~\ref{fig:cg-W} shows the error calculated as the difference of the $\zeta$ estimate and its exact value for the first 15 iterations of the Inverse Power method using the problem size W. The error decreases over successive iterations until reaching a minimum value that cannot be further decreased: \code{float} and \code{posit32} reach a minimum error value of 2.21E-4 and 1.25E-5 respectively. Quire32 has the same $\zeta$ error value as \code{posit32} and for this reason \code{posit32} and  Quire32 lines are superimposed. The CG benchmark using \code{posit32} provides a final solution that is one more digit ($\log_{10} (2.21E-4/1.25E-5) = 1.25$) accurate than the CG benchmark using \code{float}.

The right panel of Fig.~\ref{fig:cg-W} shows the norm of the residual $r = Az - x$ for the first 150 iterations. Since each inverse power iteration consists of 25 iterations, iteration 150 of CG refers to the sixth inverse power iteration. At the last CG iteration, the norm of the residual reaches a minimum value of 7.35E-7, 6.99E-8 and 2.50E-8 for \code{float}, \code{posit32} and \code{quire32} implementations respectively.  We note that in the case of error calculated as residual norm (right panel of Fig.~\ref{fig:cg-W}), the error is different for \code{posit32} and \code{quire32} and it is close in value to the machine epsilon for \code{float} and \code{posit32}: 1.19E-7 and 7.45E-9.

{\bf MG: Multigrid.} The NPB MG kernel implements the V-cycle multigrid algorithm to compute the solution of a discrete Poisson problem ($\nabla^2u = v$) on a 3D Cartesian grid. After each iteration, the $L_2$ norm of residual $r = v - \nabla^2u$ is computed for measuring the error.

\begin{figure}[h]
	\begin{center}
		\includegraphics[width=\linewidth]{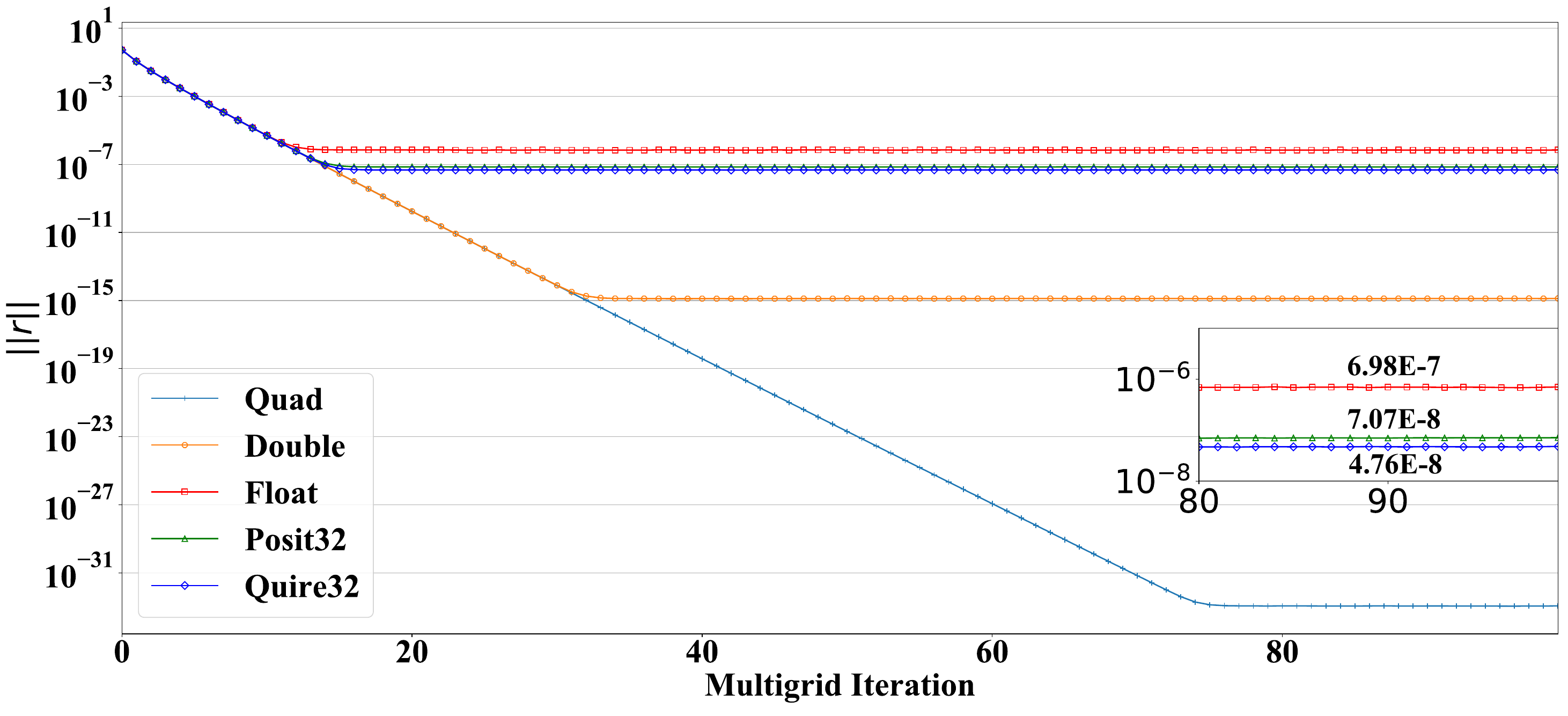}
		\caption{MG: Norm of residual over MG iterations for W class problem. \code{Posit32} and \code{quire32} implementations result in lower error when compared to \code{float} implementation. The zoom-in view shows the fine difference between \code{float}, \code{posit32} and \code{quire32} when precision improvements cannot be made with more iterations.}
		\label{fig:mg-W}
	\end{center}
\end{figure}

As for the CG benchmark, the norm of the residual in the MG application decreases during the iterations until it cannot be further reduced as shown in Fig.~\ref{fig:mg-W}. The norm of the residual for the \code{float}, \code{posit32} and \code{quire32} MG implementations at their last iteration are 6.98E-7, 7.07E-8 and 4.76E-8 respectively. These values are close to the machine $\epsilon$ values for the different floating-point formats. Also for MG, the \code{posit32} implementation provides a final solution that is one digit more accurate than the results that have been obtained with \code{float}. The \code{quire32} implementation is roughly 1.16 digit more accurate than the float implementation.

\begin{figure}[h!]
	\begin{center}
		\includegraphics[width=\linewidth]{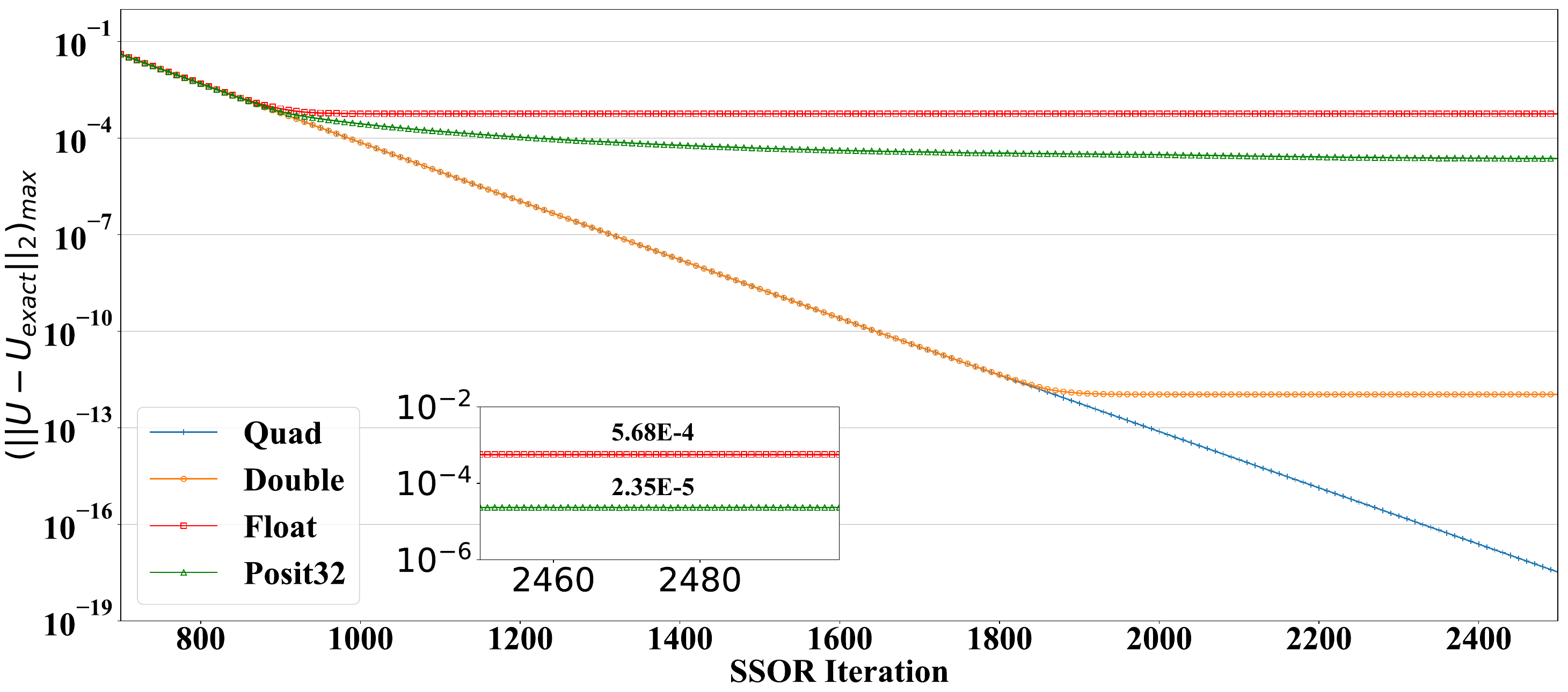}
		\caption{LU: The maximum error norm over the five PDEs from iteration 700 to 2499 for LU. The error decreases reaching a minimum error. The LU \code{posit32} implementation leads to a lower error with respect to the LU \code{float} implementation. }
		\label{fig:lu-W}
	\end{center}
\end{figure}

{\bf LU: Lower-Upper Decomposition Solver.} LU is a CFD pseudo-application solving the Navier-Stokes system comprising a system of 5 PDEs using the Symmetric Successive Over-Relaxation (SSOR) technique. In this case, we compute the error as the difference between the estimate and the analytical solution and taking its norm over the five PDEs. This error over several iterations is shown in Fig.\ref{fig:lu-W}.
The LU \code{float} implementation reaches a minimum error of 5.68E-4 while the LU \code{posit32} implementation reaches an error of 2.35E-5. The LU \code{posit32} implementation is 1.38 digit more accurate than the LU \code{float} implementation. 

\begin{figure}[h!]
	\begin{center}
		\includegraphics[width=\linewidth]{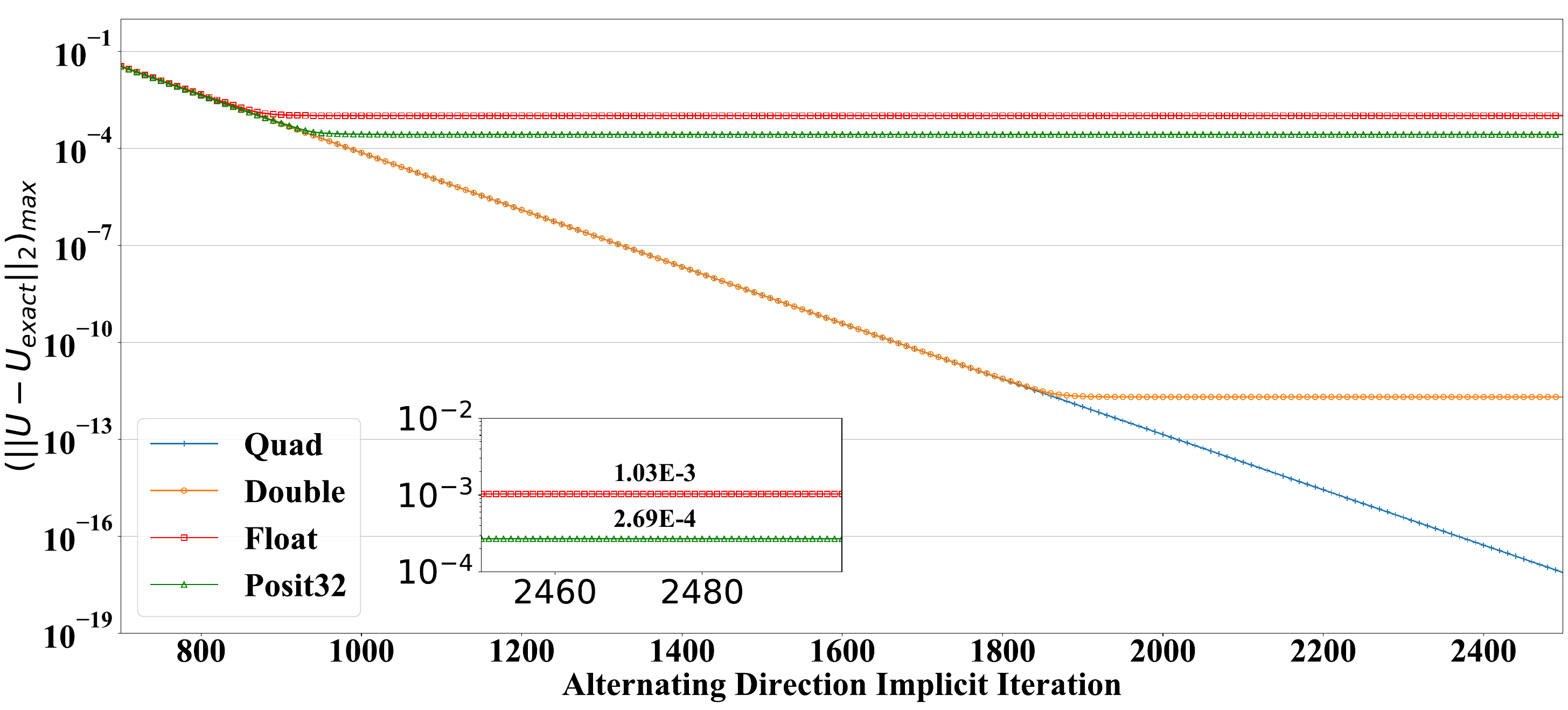}
		\caption{BT: The maximum error norm over the five PDEs at each iteration from iteration 700 to 2499 for BT. The \code{posit32} implementation results in a lower error when compared to the \code{float} implementations.}
		\label{fig:bt-W}
	\end{center}
\end{figure}

{\bf  BT: Block Tridiagonal Solver.} BT is also a CFD pseudo-application. BT employs an implicit discretization of compressible Navier-Stokes equations in 3D Cartesian geometry. The discrete solution of these equations is based on an Alternating Direction Implicit (ADI) approximate factorization. ADI factorization decouples the $x$, $y$, and $z$ directions, resulting in a linear system with block-tridiagonal of $5 \times 5$ blocks structure that is solved sequentially along each direction. We evaluate the error in the same way we evaluated the error for LU and show it in Fig.\ref{fig:bt-W}.

The \code{float} version of BT application has an error of 1.03E-3 at the last iteration while the \code{posit32} implementation has an error of 2.69E-4. The \code{posit32} implementation is 0.6 digit more accurate than the \code{float} version.

{\bf FT: Fast Fourier Transform.}  The NPB FT kernel solves a 3D Partial Differential Equation $\partial u(x,t)/\partial t = \alpha \nabla^2 u(x,t)$ in the spectral space using forward and inverse FFT. The application consists of three steps. Forward FFT, Evolve by multiplying a factor and inverse FFT. The solver computes the FFT of the state array $\tilde{u}(k,0) = FFT(u(x,0))$ at the initial step. The solution of the PDE is then advanced in the spectral space through an \emph{Evolve} process by multiplying $\tilde{u}(k,0)$ by an exponential complex factor. At each iteration, an Inverse FFT (IFFT) is computed on the result state array to move the solution from the spectral space to real space. 

\begin{figure}[h]
	\begin{center}
		\includegraphics[width=\linewidth]{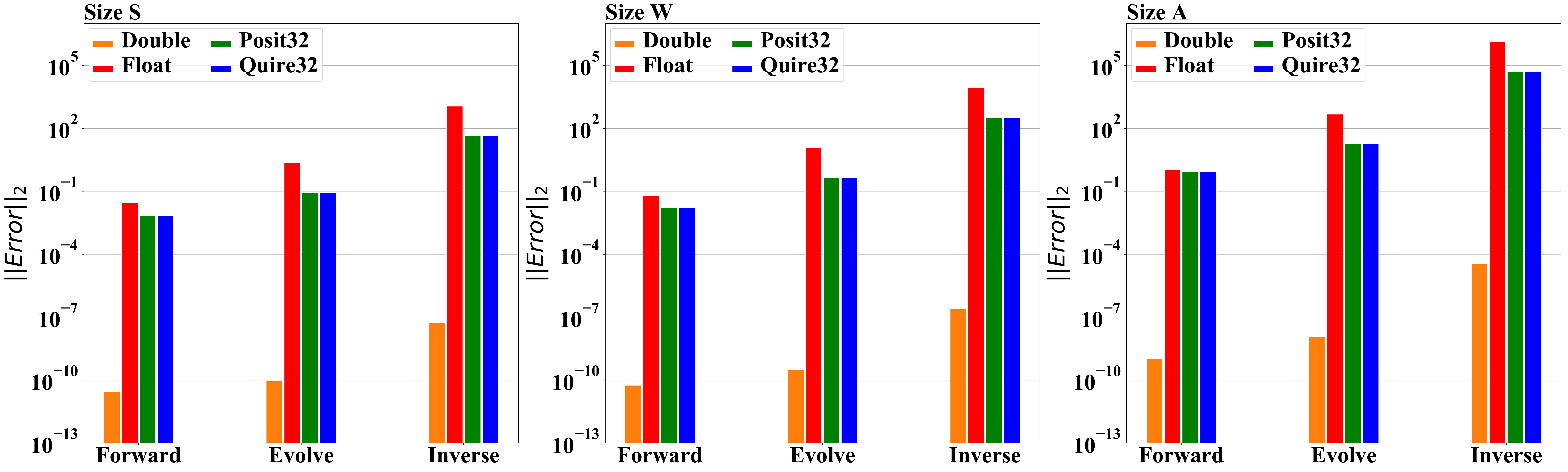}
		\caption{FT: Error norm of FT for problem classes S, W and A. The ground truth is the result for the same step computed in Quad precision. FT Float implementation displays a generally higher error when compared to the \code{posit32} implementation.}
		\label{fig:ft}
	\end{center}
\end{figure}

Fig.~\ref{fig:ft} shows that the FT \code{posit32} implementation gives a generally higher accuracy than the \code{float} implementation. Computation results using \code{quad} at each step is used as the truth. For Size W, \code{float}, \code{posit32} and \code{quire32} give 5.85E-2, 1.60E-2 and 1.59E-2 errors in the Forward step. After the Evolve step, the errors are 1.15E+1, 4.38E-1 and 4.38E-1 respectively. Finally, after the reverse FFT, the errors are 8.30E+3, 3.19E+2 and 3.18E+2 respectively. The \code{posit32} implementation is 1.4 digit more accurate than the \code{float} implementation.

{\bf Posit Performance.} In this work, we have used a software implementation of Posit floating-point format resulting in performance overhead. Table \ref{table:exec-time} presents the average execution time and standard deviation of for different implementations of NPB with size S. The execution time includes also time for the validation test and possibly I/O.  However, it gives an idea of how execution time differs between the different implementations. Quire implementations for LU and BT are not available so their execution time is not reported in the table.
\begin{table}[h]
	\setlength{\tabcolsep}{4pt}
	\renewcommand{\arraystretch}{1.1}
	\centering
	\caption{Average execution time and standard deviation of different implementation of benchmark with size S in seconds over five executions.}
	\begin{tabular}{|l|l|l|l|l|l|}
		\hline
		               & \textbf{CG}    & \textbf{FT}   & \textbf{MG}   & \textbf{LU}   & \textbf{BT}   \\ \hline
		Iter/Steps     & 30    & 6    & 100  & 100  & 100  \\ \hline
		\code{quad}    & 5.37$\pm$0.014  & 5.35$\pm$ 0.032  & 4.30$ \pm$ 0.006  & 4.52$\pm$ 0.040 & 7.96$\pm$0.065 \\ \hline
		\code{double}  & 3.16$\pm$0.006  & 0.70$\pm$ 0.002  & 0.27$\pm$ 0.003  & 0.42$\pm$ 0.011 & 0.56$\pm$ 0.003 \\ \hline
		\cellcolor[HTML]{C0C0C0}\code{float}   & \cellcolor[HTML]{C0C0C0}3.10$\pm$\cellcolor[HTML]{C0C0C0}0.053  & \cellcolor[HTML]{C0C0C0}0.67$\pm$ 0.009  & \cellcolor[HTML]{C0C0C0}0.24$\pm$ 0.003  & \cellcolor[HTML]{C0C0C0}0.43$\pm$ 0.003 & \cellcolor[HTML]{C0C0C0}0.54$\pm$ 0.001 \\ \hline
		\cellcolor[HTML]{C0C0C0}\code{posit32} & \cellcolor[HTML]{C0C0C0}12.78$\pm$ 0.070 & \cellcolor[HTML]{C0C0C0}5.73$\pm$ 0.038  & \cellcolor[HTML]{C0C0C0}4.59$\pm$ 0.018  & \cellcolor[HTML]{C0C0C0}5.74$\pm$ 0.023 & \cellcolor[HTML]{C0C0C0}7.81$\pm$ 0.040 \\ \hline
		\code{quire32} & 12.29$\pm$ 0.143 & 7.94$\pm$ 0.009  & 5.41$\pm$ 0.016 & N/A  & N/A  \\ \hline
	\end{tabular}
	\label{table:exec-time}
\end{table}
The use of Posit and Quire results in lower performance in terms of execution time and most cases are similar to results from Quad, which is also software based. For example, our CG benchmark in Posit executes $4\times$ slower than its counterpart using IEEE Float.
\section{Related Work}\label{sec:related-work}
Taper precision floating-point format has a long history of development. The concept was originally proposed in the 1970s, where a larger size exponent can be used to represent a larger number range with diminishing accuracy due to reduction in fraction bits. Universal number (unum) is another floating-point format that embraces a similar concept. Unum follows the same structure of IEEE 754 floating-point format but specifies the sizes of each component by encoding them at the end of the binary string respectively. The sizes of each component vary automatically~\cite{gustafson2015end}. Among other features, it specifies a special "ubit" which indicates if the number represented is exact or lies in an open number interval. Another radically different representation of fraction number is fix point representation. Fix point represents a real number as an integer and the fractional part can be identified through shifting of digit. An integer can be conceived as a subset of fix point number system where shifting is zero.

Standardization of Posit is currently underway\footnote{\url{https://posithub.org/docs/posit_standard.pdf}} and several implementations are available. Among those, the most complete implementation is \emph{SoftPosit}. \emph{SoftPosit}\footnote{\url{https://gitlab.com/cerlane/SoftPosit}} is implemented as a C library and is endorsed by the Next Generation Arithmetic (NGA) team\footnote{\url{https://posithub.org/docs/PDS/PositEffortsSurvey.html}}.
\section{Discussion \& Conclusion}\label{sec:conclusion}

In this work, we assessed the precision optimization in HPC applications using an emerging precision format called Posit. Our results showed that typical HPC kernels as in the NPB suite could improve their accuracy by using 32-bit Posit instead of 32-bit IEEE float. All tested kernels in our Posit NPB suite achieved higher precision, ranging from 0.6  to 1.4 decimal digit, compared to the IEEE Float baseline. However, a major obstacle that hinders the adoption of Posit is the overhead of software implementation. Our Posit NPB suite quantifies 4-19$\times$ overhead that of IEEE formats. This high overhead can be partially attributed to the operator overloading in C++, but more importantly, to the lack of hardware support. For the adoption of Posit by HPC applications, hardware implementations are necessary to achieve acceptable performance~\cite{8425396}. Overall, our results indicate Posit as a promising drop-in replacement for IEEE Float in HPC applications for precision optimization.

\section*{Acknowledgments}
{Funding for the work is received from the European Commission H2020 program, Grant Agreement No. 801039 (EPiGRAM-HS). LLNL release: LLNL-PROC-779741-DRAFT}
\vspace{-0.2cm}
%
%
%
%

\bibliographystyle{splncs04}
\bibliography{main}

\begin{thebibliography}{10}
\providecommand{\url}[1]{\texttt{#1}}
\providecommand{\urlprefix}{URL }
\providecommand{\doi}[1]{https://doi.org/#1}

\bibitem{nas}
{An unofficial C version of the NAS parallel Benchmarks OpenMP 3.0} (2014),
  \url{https://github.com/benchmark-subsetting/NPB3.0-omp-C}

\bibitem{anzt2019adaptive}
Anzt, H., Dongarra, J., Flegar, G., Higham, N.J., Quintana-Ort{\'\i}, E.S.:
  Adaptive precision in block-jacobi preconditioning for iterative sparse
  linear system solvers. Concurrency and Computation: Practice and Experience
  \textbf{31}(6),  e4460 (2019)

\bibitem{bailey1991parallel}
Bailey, D.H., Barszcz, E., Barton, J.T., Browning, D.S., Carter, R.L., Dagum,
  L., Fatoohi, R.A., Frederickson, P.O., Lasinski, T.A., Schreiber, R.S.,
  et~al.: {The NAS parallel benchmarks}. The International Journal of
  Supercomputing Applications  \textbf{5}(3),  63--73 (1991)

\bibitem{dongarra2011international}
Dongarra, J., Beckman, P., Moore, T., Aerts, P., Aloisio, G., Andre, J.C.,
  Barkai, D., Berthou, J.Y., Boku, T., Braunschweig, B., et~al.: The
  international exascale software project roadmap. International Journal of
  High Performance Computing Applications  \textbf{25}(1),  3--60 (2011)

\bibitem{gustafson2015end}
Gustafson, J.L.: {The End of Error: Unum Computing}. Chapman and Hall/CRC
  (2015)

\bibitem{gustafson2017beating}
Gustafson, J.L., Yonemoto, I.T.: Beating floating point at its own game: Posit
  arithmetic. Supercomputing Frontiers and Innovations  \textbf{4}(2),  71--86
  (2017)

\bibitem{johnson2018rethinking}
Johnson, J.: {Rethinking floating point for deep learning}. arXiv preprint
  arXiv:1811.01721  (2018)

\bibitem{Lindstrom:2018:UCR:3190339.3190344}
Lindstrom, P., Lloyd, S., Hittinger, J.: {Universal Coding of the Reals:
  Alternatives to IEEE Floating Point}. In: Proceedings of the Conference for
  Next Generation Arithmetic (2018). \doi{10.1145/3190339.3190344}

\bibitem{markidis2018nvidia}
Markidis, S., Chien, S.W.D., Laure, E., Peng, I.B., Vetter, J.S.: Nvidia tensor
  core programmability, performance \& precision. In: 2018 IEEE International
  Parallel and Distributed Processing Symposium Workshops (IPDPSW). pp.
  522--531. IEEE (2018)

\bibitem{Menon:2018:AAD:3291656.3291720}
Menon, H., Lam, M.O., Osei-Kuffuor, D., Schordan, M., Lloyd, S., Mohror, K.,
  Hittinger, J.: {ADAPT: Algorithmic Differentiation Applied to Floating-point
  Precision Tuning}. In: Proceedings of the International Conference for High
  Performance Computing, Networking, Storage, and Analysis (2018)

\bibitem{8425396}
{Podobas}, A., {Matsuoka}, S.: {Hardware Implementation of POSITs and Their
  Application in FPGAs}. In: 2018 IEEE International Parallel and Distributed
  Processing Symposium Workshops (IPDPSW). pp. 138--145 (May 2018).
  \doi{10.1109/IPDPSW.2018.00029}

\end{thebibliography}

\end{document}